\documentclass[runningheads,citeauthoryear]{apinv}
\usepackage{epsfig,cite,graphics}
\usepackage{marvosym} 
\usepackage[utf8]{inputenc}

\begin{document}

\title{New Results from Long-time Photometric Study \\ of UX Orionis Star GM Cephei}
\titlerunning{New Results from Long-time Photometric Study of UX Orionis Star GM Cepeus}
\author{Asen Mutafov\inst{1}, Evgeni Semkov\inst{1}, Stoyanka Peneva\inst{1}, Sunay Ibryamov\inst{2}}
\authorrunning{A. Mutafov et al.}
\tocauthor{Rumen Bogdanovski, Nobody Else} 
\institute{ Institute of Astronomy and NAO, Bulgarian Academy of Sciences, BG-1784, Sofia
	\and Shumen University, 115, Universitetska St, BG-9700, Shumen
	\newline
	\email{amutafov@astro.bas.bg}    }
\papertype{Submitted on xx.xx.xxxx; Accepted on xx.xx.xxxx}	
\maketitle

\begin{abstract}
New results from long-term optical photometric observations of the pre-main sequence star GM Cep from UX Orionis type are reported. During ongoing photometric monitoring of the GM Cep four deep minimums in brightness are observed. The collected multicolour photometric data shows the typical of UXor variables colour reversal during the minimums in brightness. Recent $BVRI$ photometric observations of GM Cep have been collected from November 2014 to October 2020.

\end{abstract}

\section*{1. Introduction}

The pre-main sequence star (PMS) GM Cep is situated in the field of the young open cluster Trumpler 37 ($\sim$4 Myr old and at a distance of 870 pc, Contreras et al. \cite{AB}).  It is likely to be a member of the cluster, Marschall and van Altena  \cite{CD}, Sicilia-Aguilar et al. \cite{EE}. 

According to the study of Sicilia-Aguilar et al. \cite{EF} GM Cep has a solar-like mass ($\it M$ $\sim$ $2.1 M_\odot$), with radius that ranges between 3 and 6 $R_\odot$. It is from G7V-K0V spectral type and with a strong IR excesses that can be explained  by the presence of a very luminous and massive circumstellar disk. In the study is found out that the  H$\alpha$ emission line in the spectrum of GM Cep has a strong P Cyg profile. It is also observed that the equivalent width of the line varies significantly from 6$\AA$ to 19$\AA$ and that the variable accretion rate is up to $\sim$ 10$^{-6}$ $M_\odot$/year.  

A long-term photometric study of GM Cep was made by Xiao et al. \cite{A} in the period of several decades. The long-term B and V light curves of the star are made by using the photographic plate archives from Harvard College Observatory and from Sonneberg Observatory. The results suggest that the light curves of GM Cep seem to be dominated by dips superposed on the quiescence state and that there is a lack of fast rises in brightness typical of EXor variables. In their study, Xiao et al.  \cite{A} have not found evidence for periodicity of observed dips in brightness.

The results from the $BVRI$ photometric observations of GM Cep collected in the period 2008 June - 2014 August and reported in our previous  studies (Semkov and Peneva\cite{B} and Semkov et al. \cite{D}) show very strong photometric variability of the star. In this period we have registered five deep minimums in brightness in the light curve of GM Cep. On the basis of these observations back then we concluded that the variability of GM Cep is dominated by fading events rather than by bursting events. The collected multicolour photometric data shows the effect of a colour reversal at the deep minimum of brightness, which is evidence of variable extinction from the circumstellar environment, typical of UXor variables.

In the study of Chen et al. \cite{C} while carrying out an intensive $BVR$ photometric monitoring of GM Cep during the period 2009-2011 they confirm the UXor nature of its variability and suggest an early stage of planetesimal formation in the star environment. A periodicity of about 300 days of the observed deep declines in brightness is suggested by Chen and Hu \cite{F}.

The multicolour observations give us the opportunity to clarify the mechanism of the brightness
variations.

\section*{2. Observations}
The CCD observations of GM Cep cover the period from June 2008 to October 2020. They were performed in two observatories with four telescopes: Rozhen National Astronomical Observatory (Bulgaria) with its 2-metre RCC, 50/70-cm Schmidt and 60-cm Cassegrain telescopes along with Skinakas Observatory of the University of Crete (Greece) with the 1.3-m Ritchey-Cr\'{e}tien telescope. Five different types of CCD cameras were used during the observations. Their technical characteristics and optical specifications are given in Table 1. 

We used the published in the work of Semkov and Peneva \cite{B} fifteen stars in the field around GM Cep for reference. A standard set of Johnson-Cousins' filters were used for all the frames that were taken. Twilight flat fields in each filter were obtained each clear evening. All frames obtained with the ANDOR and Vers Array cameras are bias subtracted and flat fielded. CCD frames obtained with the FLI PL16803 and FLI PL09000 cameras are dark subtracted and flat fielded. Using IDL DAOPHOT routines an aperture photometry was performed. In order to obtain comparable results with our previous obserations we used the same aperture to analyse all the data. It was chosen as 6 arcsec in radius, while the background annulus was from 10 to 15 arcsec.

   \begin{table*}
   \small
 \begin{center}
   \caption[]{CCD cameras and optical specifications}
   \begin{tabular}{llllllllllll}

            \hline
            \noalign{\smallskip}

 Telescope  & CCD Camera Type& Size & Field  &Pixel&Scale      & RON        &Gain \\
&&&&size&&&\\
 &                &(px)      &(arcmin)&($\mu$m)   &(''/px)&($e^-$rms)&($e^-$/ADU)\\  
            \noalign{\smallskip}
            \hline
            \noalign{\smallskip}
2m RCC& Vers Array 1300B&1340x1300& 5.8x5.6&20.0&0.26&2.00&1.0\\
2m RCC& ANDOR iKon-L&2048x2048& 6.0x6.0&13.5&0.17&6.90&1.1\\
Schmidt& FLI PL16803&4096x4096& 73.8x73.8&9.0&1.08&9.00&1.0\\
60cm Cass& FLI PL9000&3056x3056& 16.8x16.8&12.0&0.33&8.50&1.0\\
1.3m RC&ANDOR DZ436-BV&2048x2048& 9.6x9.6&13.5&0.28&8.14&2.7\\

\noalign{\smallskip}
            \hline

         \end{tabular}
  \end{center}
   \end{table*}

 \section*{3. Results}
In the current paper the data from the multicolour photometric observations of GM Cep is presented for the period from August 2014 to October 2020. They are a continuation of the observations that were begun by us in June 2008. Previous data is published in the work of Semkov and Peneva \cite{B} for the period June 2008 - February 2011 and in the work of Semkov et al. \cite{D} for the period April 2011 - August 2014.

The new data in the period from November 2014 to October 2020 is shown in Table 2.
The columns provide the Julian date (JD) of observation, $BVRI$ magnitudes, and the telescope used. In the column Telescope the abbreviation 2-m denotes the 2-m Ritchey-Chr\'{e}tien-Coud\'{e}, Schmidt - the 50/70-cm Schmidt and the 1.3-m Ritchey-Cr\'{e}tien telescope. The values of the instrumental errors are in the range 0.$^{m}$01-0.$^{m}$05 (for B), 0.$^{m}$01-0.$^{m}$03 (for V) and  0.$^{m}$01-0.$^{m}$02 (for R and I) (Semkov and Peneva \cite{B}).

As presented by the graphics in Figure 1 the new photometric data shows continued strong brightness variability of GM Cep. The same brightness variability was also registered in the previous studies Sicilia-Aguilar et al. \cite{EF}, Xiao  et al. \cite{A}, Semkov and Peneva \cite{B}, Chen et al. \cite{C}, Semkov et al. \cite{D}, Huang et al. \cite{E}. 

In the time scale of days and months outside the deep minimums GM Cep also shows significant brightness variations. The summarized results of over 12 years period of observations show very strong photometric variability. We have registered new four deep minimums in brightness in the light curve of GM Cep: August 2015, January 2017, November 2017, August 2020.

In Figure 2 are shown respectively colour-magnitude diagrams $V$/$B-V$, $V$/$V-R$ and $V$/$V-I$. The collected multicolour photometric data shows the typical of UXor colour reversal during the minimums in brightness. This coincides with the model of blurring of the dust-like material. The observed reverse of the colour is caused by the diffused light from small dust grains. This is a typical characteristic of PMS stars from the Uxor type. From visual inspection for every of the colour diagrams such point of reversal is observed at different star brightness: in the  $V$/$B-V$ diagram, the point of reversal is observed at $V$ about 14.0 mag, in the $V$/$V-R$ diagram at $V$ about 14.5 mag and in the $V$/$V-I$ diagram at $V$ about 14.6 mag.

Usually, when in the line of sight there are clusters of dust the star becomes redder. But during a maximum eclipse the blue part of the diffused light in the observed light becomes significant and the star colour becomes bluer.


\begin{figure}[!htb]
  \begin{center}
    \centering{\epsfig{file=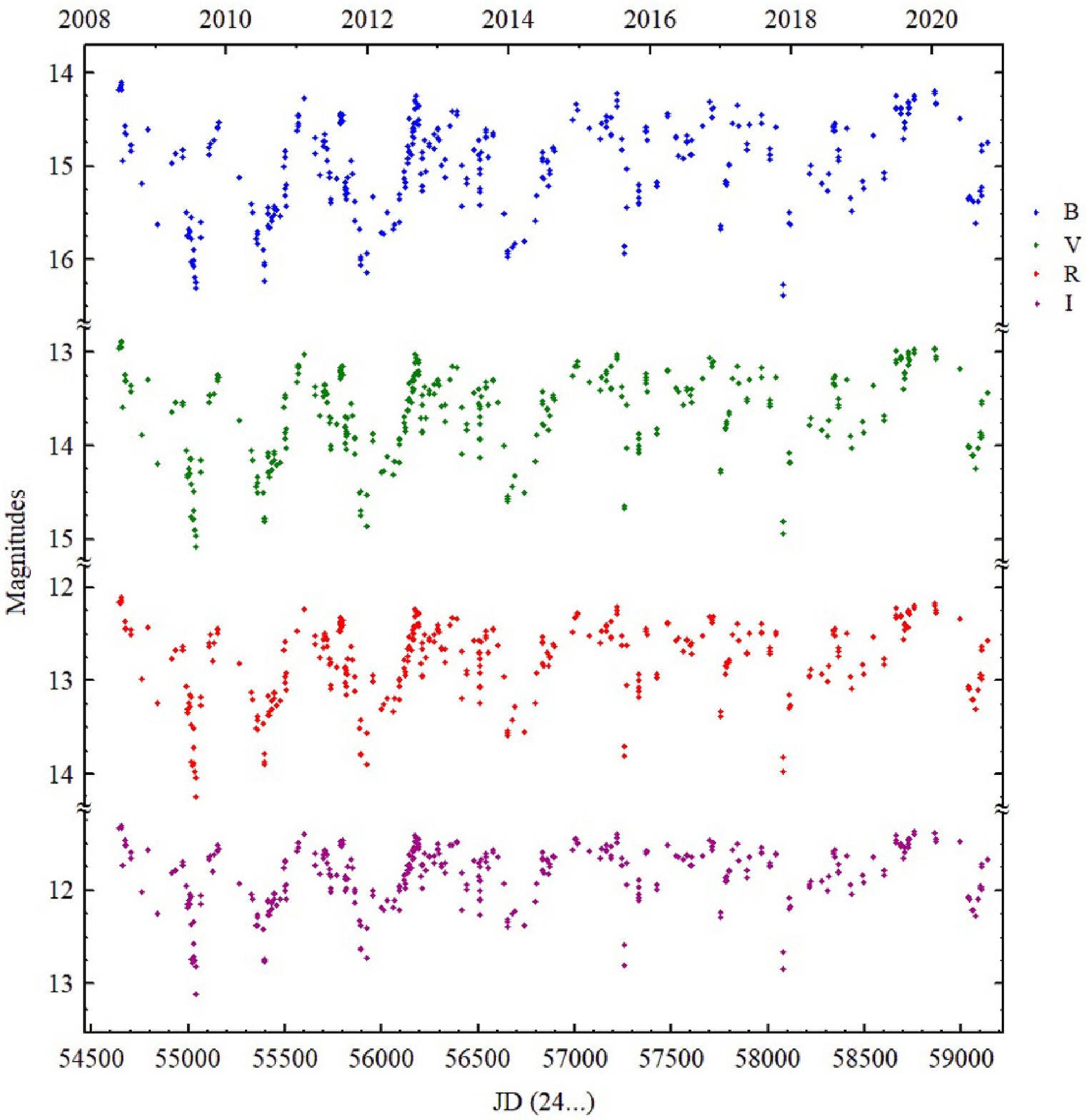, width=1.0\textwidth}}
    \caption[]{BVRI light curves of GM Cep for the whole period of our photometric monitoring (2008 - 2020).}
    \label{countryshape}
  \end{center}
\end{figure}

\begin{figure}[!htb]
  \begin{center}
    \centering{\epsfig{file=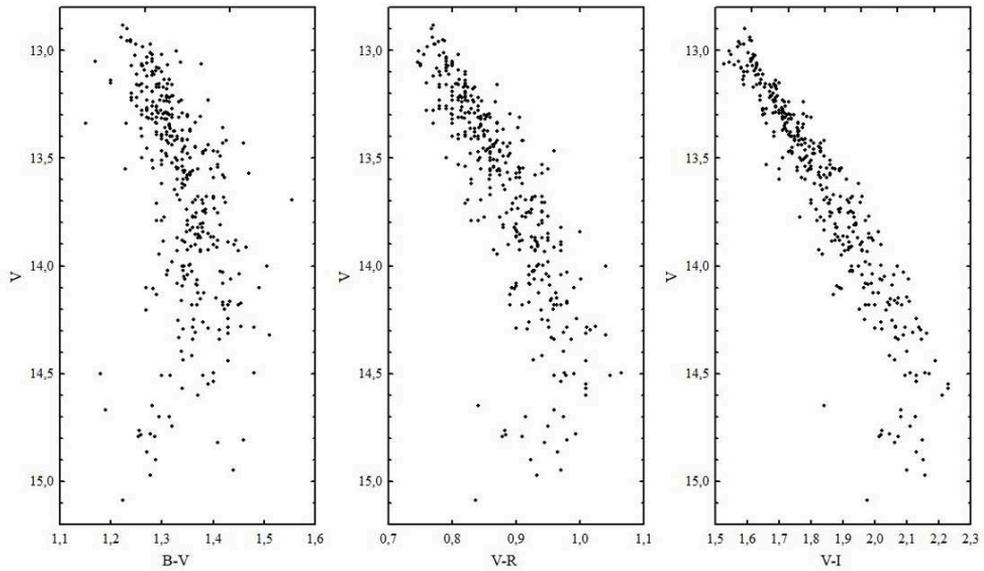, width=1.0\textwidth}}
    \caption[]{The colour-magnitude diagrams $V$/$B-V$, $V$/$V-R$ and $V$/$V-I$ of GM Cep in the period of observations June 2008 - October 2020.}
    \label{countryshape}
  \end{center}
\end{figure}

 \begin{table*}
   \small
 \begin{center}
   \caption[]{$BVRI$ photometric observations of GM Cep}
   \begin{tabular}{llllllllllll}
            \hline
            \noalign{\smallskip}
      JD (24...) & $\it B$  & $\it V$ & $\it R$ & $\it I$  & Tel& JD (24...) & $\it B$  & $\it V$ & $\it R$ & $\it I$  &Tel\\
            \noalign{\smallskip}
            \hline
            \noalign{\smallskip}

56988.217	&	14.51	&	13.26	&	12.48	&	11.57	&	Sch	&	57969.371	&	14.46	&	13.17	&	12.39	&	11.55	&	Sch	\\
57005.259	&	-	    &	13.15	&	12.33	&	11.45	&	Sch	&	58011.342	&	14.81	&	13.51	&	12.65	&	11.71	&	Sch	\\
57006.315	&	14.34	&	13.14	&	12.32	&	11.45	&	Sch	&	58012.352	&	14.93	&	13.58	&	12.72	&	11.74	&	Sch	\\
57016.251	&	14.41	&	13.16	&	12.29	&	11.49	&	2-m	    &	58013.348	&	14.88	&	13.55	&	12.69	&	11.75	&	Sch	\\
57017.202	&	14.40	&	13.10	&	12.28	&	11.51	&	2-m	    &	58039.325	&	-	    &	-	    &	12.51	&	11.61	&	Sch	\\
57074.579	&	14.60	&	13.32	&	12.52	&	11.59	&	Sch	&	58043.304	&	14.59	&	13.27	&	12.48	&	11.62	&	Sch	\\
57136.511	&	14.71	&	13.41	&	12.60	&	11.66	&	Sch	&	58080.294	&	16.27	&	14.81	&	13.83	&	12.66	&	Sch	\\
57138.450	&	14.54	&	13.27	&	12.47	&	11.56	&	Sch	&	58081.311	&	16.39	&	14.95	&	13.98	&	12.85	&	Sch	\\
57162.406	&	14.47	&	13.23	&	12.42	&	11.52	&	Sch	&	58109.346	&	15.61	&	14.18	&	13.29	&	12.20	&	Sch	\\
57164.479	&	14.58	&	13.31	&	12.47	&	11.59	&	Sch	&	58113.284	&	15.50	&	14.08	&	13.16	&	12.09	&	Sch	\\
57167.433	&	14.52	&	13.21	&	12.41	&	11.57	&	2-m	    &	58114.287	&	15.63	&	14.18	&	13.27	&	12.18	&	Sch	\\
57186.480	&	14.66	&	13.40	&	12.55	&	11.64	&	Sch	&	58217.551	&	15.08	&	13.79	&	12.96	&	11.97	&	Sch	\\
57187.498	&	14.68	&	13.39	&	12.54	&	11.66	&	2-m	    &	58218.535	&	15.09	&	13.79	&	12.95	&	11.95	&	Sch	\\
57190.410	&	14.48	&	13.16	&	12.37	&	11.53	&	2-m	    &	58220.493	&	15.00	&	13.71	&	12.89	&	11.91	&	Sch	\\
57220.402	&	14.30	&	13.02	&	12.21	&	11.40	&	Sch	&	58278.436	&	15.19	&	13.84	&	12.94	&	11.91	&	Sch	\\
57221.459	&	14.22	&	13.05	&	12.25	&	11.44	&	Sch	&	58312.365	&	15.27	&	13.90	&	13.02	&	12.00	&	Sch	\\
57223.437	&	14.37	&	13.08	&	12.29	&	11.50	&	2-m	    &	58316.323	&	15.09	&	13.74	&	12.85	&	11.85	&	Sch	\\
57246.395	&	14.83	&	13.48	&	12.62	&	11.74	&	1.3-m  	&	58340.393	&	14.62	&	13.33	&	12.51	&	11.61	&	Sch	\\
57247.411	&	14.71	&	13.37	&	12.52	&	11.66	&	1.3-m	  &	58342.379	&	14.58	&	13.28	&	12.46	&	11.58	&	Sch	\\
57259.363	&	15.86	&	14.67	&	13.71	&	12.59	&	Sch	&	58343.391	&	14.58	&	13.29	&	12.47	&	11.59	&	Sch	\\
57260.404	&	15.93	&	14.65	&	13.81	&	12.81	&	Sch	&	58344.388	&	14.55	&	13.26	&	12.45	&	11.57	&	Sch	\\
57269.401	&	15.45	&	14.03	&	13.05	&	11.94	&	Sch	&	58346.344	&	14.62	&	13.36	&	12.53	&	11.64	&	2-m	\\
57271.468	&	15.04	&	13.57	&	12.63	&	11.71	&	2-m	    &	58363.360	&	14.94	&	13.60	&	12.74	&	11.82	&	Sch	\\
57330.255	&	15.20	&	13.87	&	12.93	&	11.89	&	Sch	&	58364.323	&	14.83	&	13.50	&	12.66	&	11.73	&	Sch	\\
57331.274	&	15.26	&	13.93	&	13.00	&	11.97	&	Sch	&	58365.533	&	14.90	&	13.56	&	12.69	&	11.80	&	2-m	\\
57332.264	&	15.34	&	14.00	&	13.08	&	12.05	&	Sch	&	58409.220	&	14.60	&	13.29	&	12.49	&	11.64	&	Sch	\\
57333.265	&	15.39	&	14.04	&	13.12	&	12.08	&	Sch	&	58428.174	&	15.34	&	13.90	&	12.97	&	11.95	&	Sch	\\
57334.244	&	15.41	&	14.08	&	13.18	&	12.11	&	Sch	&	58435.295	&	15.49	&	14.04	&	13.09	&	12.04	&	Sch	\\
57369.257	&	14.63	&	13.29	&	12.47	&	11.61	&	2-m	    &	58492.225	&	15.16	&	13.75	&	12.84	&	11.84	&	Sch	\\
57370.229	&	14.62	&	13.23	&	12.45	&	11.59	&	2-m	    &	58496.201	&	15.24	&	13.87	&	12.94	&	11.92	&	Sch	\\
57371.224	&	14.59	&	13.27	&	12.44	&	11.61	&	2-m	    &	58547.625	&	14.68	&	13.36	&	12.54	&	11.65	&	Sch	\\
57372.241	&	14.63	&	13.34	&	12.47	&	11.58	&	Sch	&	58603.481	&	15.14	&	13.74	&	12.83	&	11.84	&	Sch	\\
57374.299	&	14.73	&	13.42	&	12.51	&	11.58	&	Sch	&	58604.532	&	15.07	&	13.68	&	12.77	&	11.79	&	Sch	\\
57425.221	&	15.18	&	13.82	&	12.94	&	11.94	&	Sch	&	58665.455	&	14.25	&	12.98	&	12.22	&	11.41	&	Sch	\\
57426.226	&	15.22	&	13.88	&	12.97	&	11.99	&	Sch	&	58666.459	&	14.38	&	13.10	&	12.31	&	11.47	&	Sch	\\
57483.462	&	14.47	&	13.19	&	12.39	&	11.52	&	2-m	    &	58667.494	&	14.40	&	13.12	&	12.33	&	11.49	&	Sch	\\
57484.480	&	14.44	&	13.20	&	12.38	&	11.52	&	2-m	    &	58690.370	&	14.38	&	13.08	&	12.31	&	11.50	&	2-m	\\
57522.460	&	14.69	&	13.39	&	12.58	&	11.63	&	Sch	&	58691.450	&	14.39	&	13.06	&	12.31	&	11.51	&	2-m	\\
57523.448	&	14.70	&	13.39	&	12.57	&	11.63	&	Sch	&	58692.389	&	14.44	&	13.06	&	12.31	&	11.54	&	2-m	\\
57540.447	&	14.89	&	13.43	&	12.55	&	11.65	&	2-m	    &	58704.371	&	14.72	&	13.40	&	12.56	&	11.66	&	Sch	\\
57565.485	&	14.92	&	13.56	&	12.69	&	11.67	&	Sch	&	58705.404	&	14.60	&	13.29	&	12.46	&	11.60	&	Sch	\\
57581.424	&	14.75	&	13.41	&	12.58	&	11.64	&	Sch	&	58706.402	&	14.53	&	13.23	&	12.41	&	11.55	&	Sch	\\
57582.459	&	14.67	&	13.38	&	12.56	&	11.62	&	Sch	&	58707.476	&	14.53	&	13.21	&	12.40	&	11.54	&	Sch	\\
57583.434	&	14.74	&	13.41	&	12.57	&	11.63	&	Sch	&	58726.384	&	14.44	&	13.14	&	12.44	&	11.55	&	2-m	\\
57603.384	&	14.88	&	13.47	&	12.63	&	11.74	&	2-m	    &	58727.428	&	14.38	&	13.07	&	12.28	&	11.51	&	2-m	\\
57605.412	&	14.88	&	13.54	&	12.72	&	11.74	&	Sch	&	58728.465	&	14.33	&	13.00	&	12.26	&	11.46	&	2-m	\\
57607.392	&	14.72	&	13.40	&	12.60	&	11.65	&	Sch	&	58729.395	&	14.32	&	13.02	&	12.27	&	11.47	&	2-m	\\
57664.304	&	14.57	&	13.28	&	12.52	&	11.63	&	Sch	&	58730.464	&	14.38	&	13.09	&	12.30	&	11.46	&	Sch	\\
57698.280	&	14.31	&	13.07	&	12.32	&	11.47	&	Sch	&	58758.414	&	14.25	&	12.97	&	12.20	&	11.38	&	Sch	\\
57714.306	&	14.48	&	13.16	&	12.38	&	11.57	&	2-m	    &	58759.461	&	14.29	&	13.01	&	12.23	&	11.41	&	Sch	\\
57715.272	&	14.39	&	13.12	&	12.35	&	11.53	&	2-m	    &	58864.243	&	14.22	&	12.97	&	12.20	&	11.39	&	Sch	\\
57716.286	&	14.38	&	13.10	&	12.32	&	11.50	&	2-m	    &	58865.243	&	14.20	&	12.96	&	12.18	&	11.39	&	Sch	\\
57755.231	&	15.68	&	14.29	&	13.39	&	12.29	&	Sch	&	58869.221	&	14.33	&	13.05	&	12.26	&	11.48	&	2-m	\\
57756.243	&	15.64	&	14.26	&	13.34	&	12.24	&	Sch	&	58870.238	&	14.34	&	13.07	&	12.28	&	11.46	&	Sch	\\
57781.221	&	15.16	&	13.81	&	12.93	&	11.87	&	Sch	&	58993.457	&	14.49	&	13.18	&	12.35	&	11.49	&	Sch	\\
57782.230	&	15.19	&	13.82	&	12.85	&	11.90	&	2-m	    &	59040.398	&	15.34	&	14.00	&	13.07	&	12.07	&	Sch	\\
57784.227	&	-	    &	13.75	&	12.81	&	11.86	&	2-m	    &	59041.413	&	15.36	&	14.02	&	13.09	&	12.09	&	Sch	\\
57785.251	&	15.17	&	13.75	&	12.83	&	11.89	&	2-m    	&	59042.403	&	15.33	&	14.02	&	13.09	&	12.10	&	Sch	\\
57786.222	&	15.20	&	13.77	&	12.86	&	11.91	&	2-m	    &	59059.412	&	15.37	&	14.10	&	13.21	&	12.22	&	Sch	\\
57800.205	&	15.00	&	13.67	&	12.81	&	11.80	&	Sch	&	59060.436	&	15.39	&	14.11	&	13.21	&	12.21	&	Sch	\\
57801.208	&	14.98	&	13.64	&	12.78	&	11.79	&	Sch	&	59075.371	&	15.61	&	14.25	&	13.31	&	12.28	&	2-m	\\
57817.531	&	14.54	&	13.28	&	12.51	&	11.57	&	Sch	&	59085.314	&	15.39	&	14.03	&	13.10	&	12.10	&	Sch	\\
57845.516	&	14.35	&	13.15	&	12.39	&	11.51	&	Sch	&	59101.456	&	15.27	&	13.91	&	12.98	&	11.98	&	Sch	\\
57846.547	&	14.57	&	13.34	&	12.57	&	11.68	&	Sch	&	59102.383	&	15.27	&	13.87	&	12.94	&	11.95	&	2-m	\\
57892.498	&	14.76	&	13.50	&	12.71	&	11.79	&	Sch	&	59103.464	&	15.32	&	13.91	&	12.99	&	12.00	&	2-m	\\
57893.573	&	14.83	&	13.53	&	12.72	&	11.87	&	2-m	    &	59105.338	&	15.23	&	13.89	&	12.96	&	11.97	&	Sch	\\
57904.451	&	14.56	&	13.30	&	12.50	&	11.65	&	Sch	&	59108.400	&	14.78	&	13.55	&	12.68	&	11.76	&	Sch	\\
57967.438	&	-	    &	-	    &	12.50	&	11.60	&	Sch	&	59109.350	&	14.84	&	13.52	&	12.64	&	11.73	&	Sch	\\
57968.369	&	14.54	&	13.27	&	12.49	&	11.61	&	Sch	&	59136.280	&	14.76	&	13.44	&	12.58	&	11.67	&	Sch	\\

\noalign{\smallskip}
            \hline
         \end{tabular}
  \end{center}
   \end{table*}

\section*{4. Conclusion}
 In the time scale of days and months outside the deep minimums GM Cep shows significant brightness variations. 

The last data confirms again the presence of an “blueing effect” during the minimum of brightness and is an independent evidence that the variability of GM Cep is dominated by the variable extinction. The collected multicolour photometric data shows the typical of UXor variables colour reversal during the minimums in brightness.

We can confirm our conclusions, made in our previous paper (Semkov et al. \cite{D}), that the photometric properties of GM Cep can be explained by a superposition of highly variable accretion from the circumstellar disk onto the stellar surfice and occultation from circumstellar clumps of dust, planetesimals or from other features of the circumstellar disk.

\section*{Acknowledgements}
This work was partly supported by the Bulgarian Scientific Research Fund of the Ministry of Education and Science under the grants DN 08-1/2016, DN 18-10/2017 and DN 18-13/2017.


\end{document}